\documentclass[sigconf, noacm]{acmart} 
\AtBeginDocument{%
  }

\usepackage{enumitem}
\usepackage{amsmath}
\usepackage{booktabs}
\usepackage{array}
\usepackage{caption}
\usepackage{subcaption}
\usepackage{siunitx}
\usepackage{multirow}
\usepackage{graphicx}
\usepackage{etoolbox}

\begin{document}

\title{OneSearch: A Preliminary Exploration of the Unified End-to-End Generative Framework for E-commerce Search}

\author{Ben Chen$^{*\dagger}$, Xian Guo$^*$, Siyuan Wang$^*$, Zihan Liang$^*$, Yue Lv, Yufei Ma, Xinlong Xiao, Bowen Xue, Xuxin Zhang, Ying Yang, Huangyu Dai, Xing Xu, Tong Zhao, Mingcan Peng, Xiaoyang Zheng, Chao Wang, Qihang Zhao, Zhixin Zhai, Yang Zhao, Bochao Liu, Jingshan Lv, Xiao Liang, Yuqing Ding, Jing Chen, Chenyi Lei$^{\dagger}$, Wenwu Ou, Han Li, Kun Gai}
\thanks{*Equal Contribution.}
\thanks{$\dagger$ Corresponding author.}
\affiliation{
 \institution{Kuaishou Technology, Beijing, China}
 \country{Contact: \{chenben03@kuaishou.com, leichenyi@gmail.com\}}
}

\renewcommand{\shortauthors}{Ben Chen et al.}

\begin{abstract}
Traditional e-commerce search systems employ multi-stage cascading architectures (MCA) that progressively filter items through recall, pre-ranking, and ranking stages. While effective at balancing computational efficiency with business conversion, these systems suffer from fragmented computation and optimization objective collisions across stages, which ultimately limit their performance ceiling. To address these, we propose \textbf{OneSearch}, the first industrial-deployed end-to-end generative framework for e-commerce search. This framework introduces three key innovations: (1) a Keyword-enhanced Hierarchical Quantization Encoding (KHQE) module, to preserve both hierarchical semantics and distinctive item attributes while maintaining strong query-item relevance constraints; (2) a multi-view user behavior sequence injection strategy that constructs behavior-driven user IDs and incorporates both explicit short-term and implicit long-term sequences to model user preferences comprehensively; and (3) a Preference-Aware Reward System (PARS) featuring multi-stage supervised fine-tuning and adaptive reward-weighted ranking to capture fine-grained user preferences. Extensive offline evaluations on large-scale industry datasets demonstrate OneSearch's superior performance for high-quality recall and ranking.  The rigorous online A/B tests confirm its ability to enhance relevance in the same exposure position, achieving statistically significant improvements: +1.67\% item CTR, +2.40\% buyer, and +3.22\% order volume. Furthermore, OneSearch reduces operational expenditure by 75.40\% and improves Model FLOPs Utilization from 3.26\% to 27.32\%. The system has been successfully deployed across multiple search scenarios in Kuaishou, serving millions of users, generating tens of millions of PVs daily.
\end{abstract}



\maketitle

\begin{figure}[th]
  \centering
  \includegraphics[width=\linewidth]{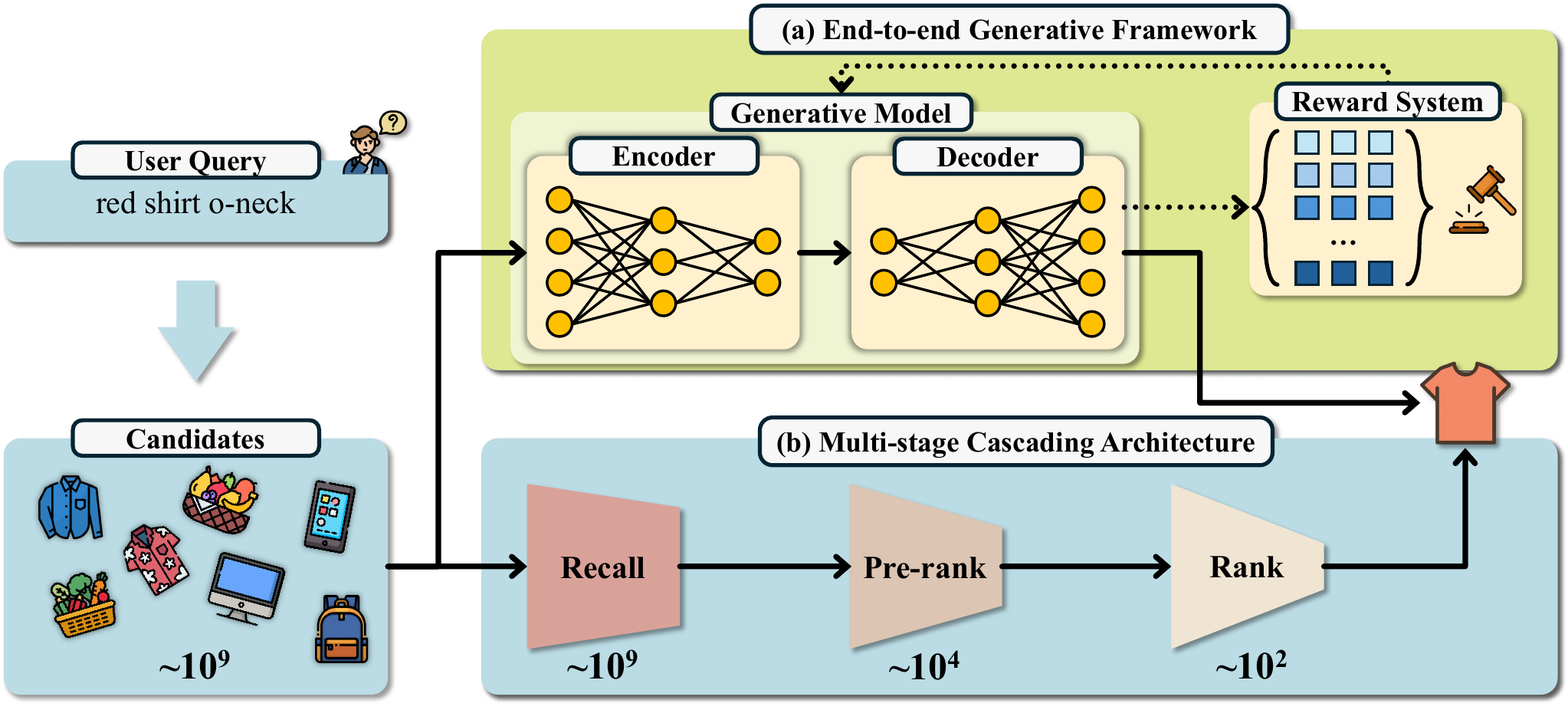}
  \caption{(a) Our proposed End-to-End generative retrieval framework, (b) the traditional multi-stage cascading architecture in E-commerce search.}
  \label{Figure1}
\end{figure}

\section{Introduction}
E-commerce search aims to retrieve items that align with the user's real intention based on their search terms, behavioral preferences, and the available inventory. To enhance user experience, these systems are typically required to identify items that satisfy both semantic and personalized criteria from hundreds of millions of candidates within one second. Consequently, traditional search systems frequently employ the Multi-stage Cascading Architecture (MCA). As depicted in Figure~\ref{Figure1}(b), MCA adheres to a coarse-to-fine paradigm, wherein a query progresses through recall, pre-ranking, and ranking stages, ultimately returning the top-selected items. The recall stage operates on the entire set of item candidates (\textasciitilde $10^9$), the pre-ranking stage narrows down the selection (\textasciitilde $10^4$), and the final ranking stage evaluates only the top candidates (\textasciitilde $10^2$).

MCA effectively balances the trade-off between system response time and sorting accuracy by progressively narrowing the pool of items at each stage. However, MCA inherently suffers from severe fragmented compute and objective collision issues \cite{2024llama3,2025onerec,2025oneloc}. Fragmented compute means most serving resources are allocated to communication and storage rather than numerical computation. Regarding objective collision, MCA potentially employs multiple strategies to meet accuracy and diversity requirements. For instance, the recall and pre-ranking stages use lightweight models that tend to retrieve all relevant items, while the ranking stage implements complex reasoning of user preferences by introducing user historical sequences, query and item statistical features. The potential discarding of intended items due to multi-layer funnel filtering, along with heterogeneous optimization objectives, reducing the performance ceiling of search systems. Furthermore, traditional MCA lacks an understanding of cold-start queries and items, resulting in limited performance in long-tail sessions \cite{2025grain, 2025onesug}.

In recent years, numerous efforts have been made to address the aforementioned issues. Some works focus on intra-stage optimization, such as EBR\cite{2020EBR} for recall, DCN\cite{2021dcn} and DSSM\cite{2013dssm} for pre-ranking, DIN\cite{2018din} and DeepFM\cite{2017deepfm} for the final ranking stage. Particularly, with the advancements in large language models (LLMs), a significant portion of research has emerged aiming to use generative models to tackle the challenges of each stage \cite{2025TBGRecall,2024hllm,2024GenR-PO,2024hstu,2025RankMixer}. Another line of work focuses on resolving the inconsistency of objectives across the full stages \cite{2023asmol,2025UniGRF,2024rankall}, striving to ensure the effective transmission of intent items through sample construction and loss design. However, these approaches still struggle to overcome the inherent limitations of MCA. For instance, (pre-)ranking stages can only process top-k items retrieved from the previous stage. If an effective item that aligns with the user's true intent is filtered out in an earlier stage, no matter how precise the subsequent models are, they cannot present this item to user.

In the past two years, a novel generative retrieval (GR) paradigm has emerged, which transforms the basic matching-based framework of MCA into generation-based approaches, addressing its inherent limitations \cite{2023tiger,2024lcrec,2025gram,2025onerec,2025onesug,2025ega,2025oneloc}. This paradigm eliminates the need for multi-stage filtering by directly inputting query or user sequence information and outputting corresponding item candidates. Tiger\cite{2023tiger} pioneered the development of end-to-end generative retrieval models for sequential recommendation, introducing the semantic IDs (SID) derived from each item's content information for efficient item representation. LC-REC\cite{2024lcrec} proposed adapting LLMs by integrating collaborative semantics for recommendation, utilizing a series of specially designed tuning tasks. Subsequently, OneRec\cite{2025onerec} was first implemented in a real industrial scenario, followed by OneSug\cite{2025onesug} for query suggestion in e-commerce search, EGA\cite{2025ega} for advertising, and OneLoc\cite{2025oneloc} for local life services.

\begin{figure}[t]
\centering
\begin{minipage}[ht]{0.32\columnwidth}
    \includegraphics[width=\columnwidth]{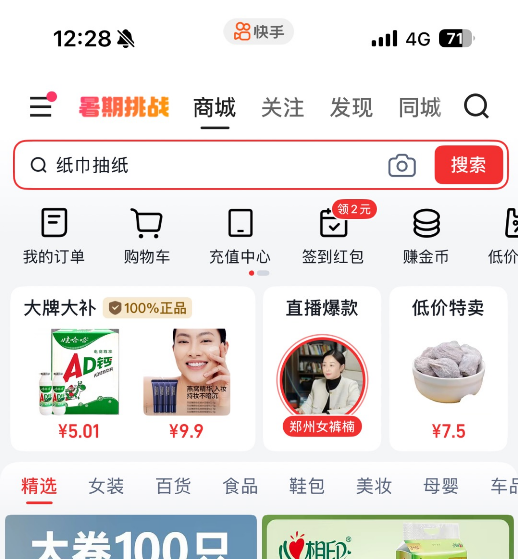}
    \subcaption{Homepage}
    \label{fig:homepage}
\end{minipage}
\begin{minipage}[ht]{0.32\columnwidth}
    \includegraphics[width=\columnwidth]{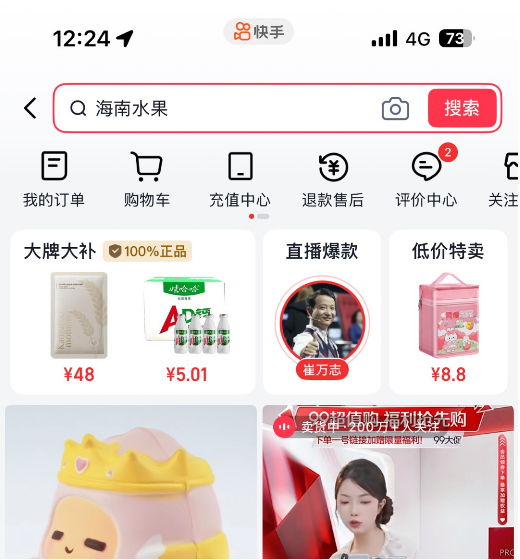}
    \subcaption{Mall}
    \label{fig:mall}
\end{minipage}
\begin{minipage}[ht]{0.32\columnwidth}
    \includegraphics[width=\columnwidth]{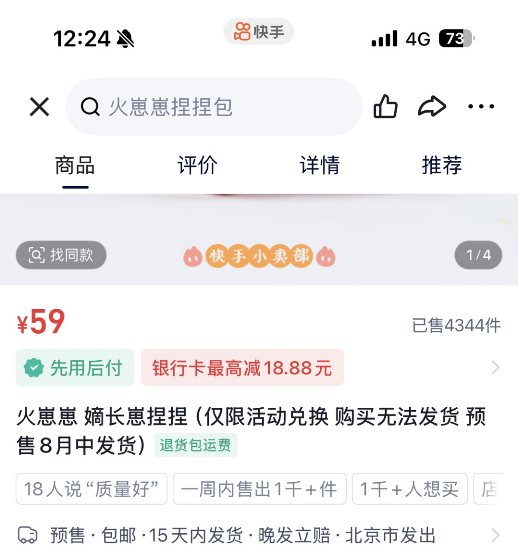}
    \subcaption{Detail Page}
    \label{fig:detailpage}
\end{minipage}
\caption{The main search entries in the Kuaishou Platform.}
\label{fig:search_entry}
\end{figure}

While for e-commerce search, these methods are not so effective due to several unique and critical challenges:
1) Item information, such as titles, keywords, and detail pages, tends to be lengthy with redundant irrelevant noise, as sellers often add unrelated keywords to increase exposure. Moreover, the semantic order in item descriptions is weak. Essential information such as brand names, attribute words, and categories often appears in the text without regard for position \cite{2024GenR-PO}. This lack of global coherence can severely mislead representation models, leading to incorrect judgments. 2) Unlike recommendations, there is a strong relevance constraint between search queries and items. Queries typically consist of 2-3 short keywords, and any mismatch in attributes can result in significant relevance issues. Although semantic ID-based GR models can construct hierarchical, learnable representations of items, they inevitably lead to the loss of core attribute representations, as they tend to learn shared information under the same SIDs. 3) Uncovering the users' latent search intents is also a core challenge. When users enter concise queries or search for a completely new category, it is crucial to effectively combine query content with user behavior profiles to infer the user's true search intent.

To address these challenges, we propose \textbf{OneSearch}, an end-to-end generative framework for e-commerce search, which includes:

1) \textbf{K}eyword-enhanced \textbf{H}ierarchical \textbf{Q}uantization \textbf{E}ncoding module. KHQE employs a keyword-enhanced semantic collaborative encoder to highlight the core attributes of items. It then uses RQ-Kmeans for hierarchical feature encoding and OPQ for unique features quantization of each item. This encoding effectively reduces interference from redundant irrelevant noise, thereby enhancing the relevance between queries and generated items.

2) \textbf{M}ulti-view \textbf{U}ser Behavior \textbf{Seq}uence (Mu-Seq) injection strategy. This strategy introduces a weighted decay click behavior sequence into the user ID to construct a distinctive user representation, then explicitly incorporates short behavior sequences in prompts to learn recent user preferences and implicitly includes long behavior sequences to model the user profile, achieving multi-view modeling of user personalized behavior.

3) \textbf{P}reference \textbf{A}ware \textbf{R}eward \textbf{S}ystem (PARS). We design a multi-stage supervised fine-tuning process for semantic alignment and personalization, followed by an adaptive reward system that leverages hierarchical user behavior signals and list-wise preference optimization. This system enables the model to learn preference differences among items while maintaining query-item relevance constraints through hybrid ranking that combines reward model guidance with direct user interaction feedback.

Extensive offline evaluations are conducted on real user search logs, and the significant performance boosts demonstrate OneSearch's effectiveness for e-commerce search. We also make multiple strict A/B testings in the KuaiShou mall search platform to verify its online effectiveness. The pure OneSearch can get comparable performance to the online MCA. By the introduction of RQ-OPQ and long behavior sequence, OneSearch can confidently improve item CTR by 1.45\%, PV CTR by 1.40\%. While applying the additional reward model selection can yield a 1.10\% increase in search pvs, 1.67\% in item CTR, 3.14\% in PV CTR, 1.78\% in PV CVR, 2.40\% in buyers volume, and 3.22\% in order volume. As a side note, we further performed an MCA testings that only included recall, pre-ranking stage, while item CTR dropped by 9.97\% and order by 39.14\%. These results effectively demonstrate the performance of OneSearch. Finally, the pure OneSearch can save about 75.40\% operational expenditure and improve the Model Flops Utilization (MFU) from 3.26\% of MCA to 27.32\%. It has been successfully deployed for the entire traffic on the detail page search, 50\% traffic on the mall search, and 20\% traffic on the homepage search platform, for further investigation. To the best of our knowledge, it is the first industrial-deployed end-to-end generative framework for e-commerce search. We hope that exploration could further pave the way for smarter GRs in Search.

The main contributions of this work are summarized as follows:
\begin{itemize}[left=0pt]
\item We propose a novel keyword-enhanced hierarchical quantization encoding, which can generate semantic IDs balanced with context features and collaborative signals for queries/items. The keyword enhancement strategy can further reduce the Interference from irrelevant noise, and strengthen the relevance constraints of GRs.
\item We devise a multi-view user behavior sequential injection strategy, with introducing sequence into the user ID representation, and the input prompt explicitly and implicitly. This approach can facilitate GRs’ reasoning about user profiles and preferences.
\item We design a preference aware reward system, which contains a multi-stage SFT process, as well as an adaptive reward model, to improve the model's personalized ranking capability.  
\item Finally, we presented OneSearch, the first industrial-deployed end2end generative framework for e-commerce search. Various offline and online A/B tests are conducted, verifying its effectiveness and efficiency for the real e-commerce search scenarios.
\end{itemize}

\section{Related Works}
\subsection{Generative Retrieval and Recommendation}
In recent years, Generative Retrieval (GR) has garnered significant attention from both academia and industry due to its remarkable performance. This emerging retrieval paradigm, which regards large-scale retrieval as sequence-to-sequence generation tasks, has outperformed traditional ANN-based models such as EBR \cite{2020EBR} and RocketQA \cite{2021rocketqa}, spurring increased exploration in the fields of search and recommendation. Notable contributions in this area include Tiger \cite{2023tiger}, DSI \cite{2022dsi}, and LC-REC \cite{2024lcrec}. Tiger \cite{2023tiger} pioneered the development of end-to-end generative retrieval models for sequential recommendation, introducing semantic IDs (SID) derived from each item's content information for efficient item representation. LC-REC \cite{2024lcrec} proposed adapting large language models (LLMs) by integrating collaborative semantics for recommendation, utilizing a series of specially designed tuning tasks.

Most GR models serve merely as supplementary recall sources within online systems, thereby overlooking these models' inherent rich semantic and powerful reasoning abilities for potential use in (pre-)ranking stages. In the area of video recommendation, OneRec \cite{2025onerec} was the first to unify recall, pre-ranking, and ranking within a single generative model. This was achieved with the assistance of session-wise generation and iterative preference alignment, resulting in substantial improvements in practical online metrics. EGA \cite{2025ega} represents a significant departure from both traditional multi-stage cascading architectures (MCA) and existing generative retrieval models by introducing a unified framework that holistically models the entire advertising pipeline. UniROM \cite{2025UniROM} employs a hybrid feature service to efficiently decouple user and advertising features, and RecFormer \cite{2023Recformer}, a variation of Transformer, captures both intra- and cross-sequence interactions.

\begin{figure}[t]
  \centering
  \includegraphics[width=\linewidth]{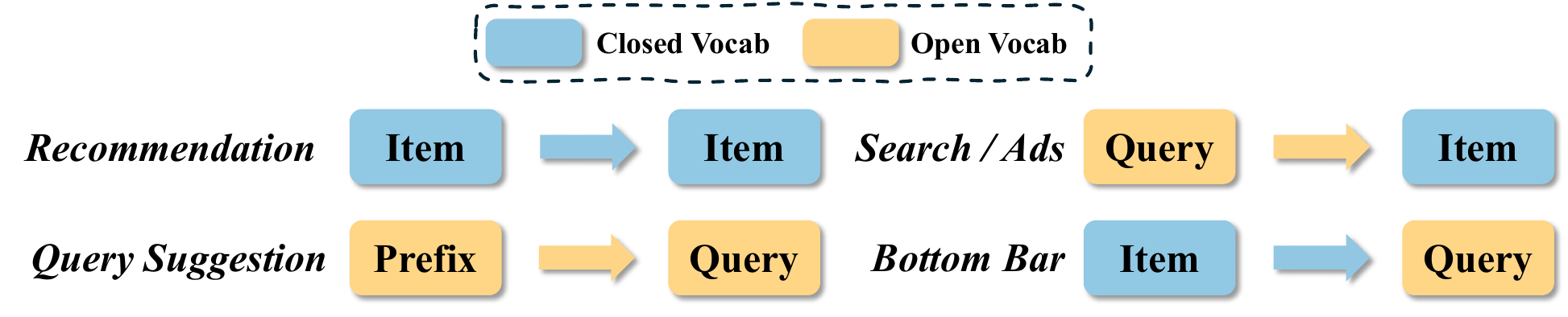}
  \caption{The input and output differences among Recommend, Search/Ads, Query Sug and Bottom Bar.}
  \label{reco_sug_sec_bar}
\end{figure}

\subsection{Generative Retrieval for Search}
These two years, most advancements in generative retrieval have been focused on recommendations. This is because search systems face three major challenges: 1) multiple and low-density item information, 2) strong relevance constraints between search queries and items, and 3) inference barriers to users' potential search intentions. Consequently, the current traditional e-commerce search systems still adopt a multi-stage cascading architecture. However, some efforts have been made to optimize current search systems using generative retrieval (GR). 

The first example is GenR-PO \cite{li2024generativeretrievalpreferenceoptimization}, which utilizes multi-span identifiers to represent raw item titles. This approach transforms the task of generating titles from queries into the task of generating multi-span identifiers from queries, thereby simplifying the generation process. Subsequently, a constrained search method is employed to identify key spans for retrieving the final item, which has proven beneficial for online recall systems. Another notable example is the Generative Retrieval and Alignment Model (GRAM) \cite{2025gram}, which performs joint training on text information from both queries and products to generate shared text identifier codes. GRAM employs a co-alignment strategy to optimize these codes for maximizing retrieval efficiency and is deployed on the JD search engine to enhance both the recall and pre-ranking stages.

Recently, we proposed OneSug \cite{2025onesug}, which incorporates a prefix2query representation enhancement module to enrich prefixes using semantically and interactively related queries to bridge content and business characteristics, an encoder-decoder generative model that unifies the query suggestion process, and a reward-weighted ranking strategy with behavior-level weights to capture fine-grained user preferences. It is the first end-to-end generative framework for e-commerce query suggestion, and has been verified to have substantial improvements in user clicks and conversion.

These GR methods demonstrate appealing performance in the realm of search, recommendation, bottom navigation, advertising, and even query suggestion. They are not suitable for e-commerce search. As illustrated in Figure~\ref{reco_sug_sec_bar}, the inputs and outputs of recommendation are the closed-vocabulary items or videos, thus the pure semantic ID tokenization is suitable for its diverse item generation. The inputs and outputs of query suggestion are the full open-vocabulary textual descriptions, so that it can directly use the transformer architecture. For the bottom bar and search engine, either the inputs or the outputs are open-vocabulary, which represents a significant departure from both OneRec and EGA.

\begin{figure*}[htbp]
    \centering
    \includegraphics[width=\textwidth]{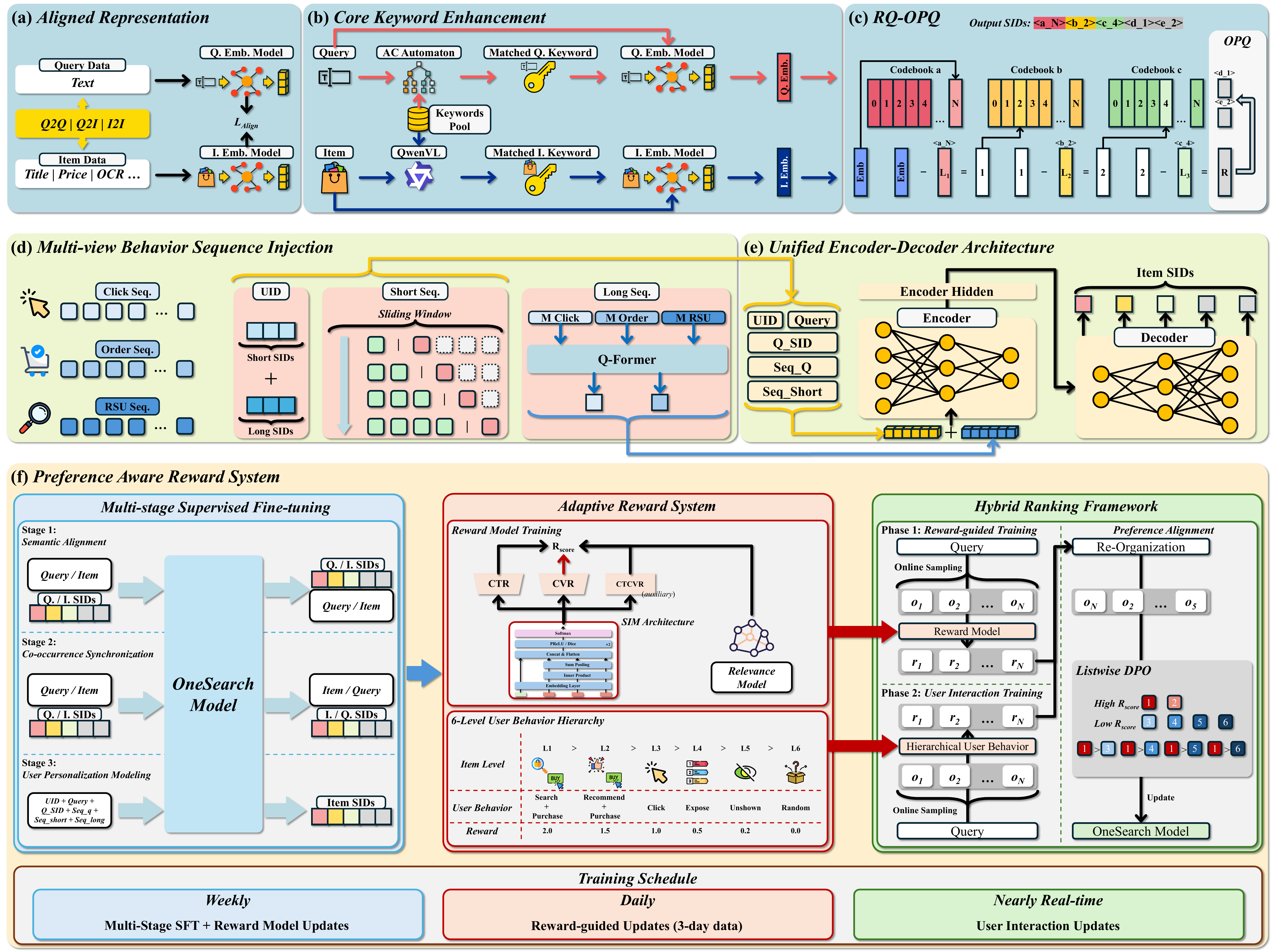}
    \caption{The Framework of OneSearch includes four parts: 1) Keyword-enhanced Hierarchical Quantization Encoding, which adopts aligned representation and core keyword to construct an elaborative hierarchical quantization tokenization schema; 2) Multi-view Behavior Sequence Injection, utilizing sequences to reason the user profiles and preferences;  3) Unified encoder-decoder Architecture that integrates produced features for generative retrieval; and 4) Preference aware reward system, which contains a multi-stage SFT procedure and an adaptvie reward system, to enhance the model's personalized ranking capability.}
    \label{figure4}
\end{figure*}

\section{Methodology}
In this section, we detail the proposed OneSearch, an end-to-end framework for e-commerce search, in four parts. We elaborate on the keyword-enhanced hierarchical quantization encoding in  \S~\ref{KHQE}, then we introduce the strategy of multi-view user behavior sequence injection in \S~\ref{Mu_seq} and unify the encoder-decoder architecture in \S~\ref{Architecture}. In \S~\ref{PARS}, a preference aware reward system is proposed, which includes a multi-stage supervised finetuning process, as well as an adaptive reward system for personalized ranking learning. The framework of OneSearch is illustrated in Figure~\ref{figure4}.

\subsection{Hierarchical Quantization Encoding}
\label{KHQE}
Encoding items into Semantic IDs (SIDs) is crucial for the success of generative retrieval models. This process converts continuous semantic representations into discrete ID sequences using a coarse-to-fine quantization approach, ensuring that items with the same SID share the same information \cite{2023tiger,2025onerec,2025GRID}. However, common quantization methods (e.g., VQ-VAE \cite{2018vqvae}, RQ-VAE \cite{2022rqvae}, or RQ K-means \cite{luo2024qarm}) tend to tokenize shared signals among items using a reduced and fixed vocabulary, which results in the loss of distinctive attributes for each item. This information loss causes many similar but not identical items sharing the same SIDs, while lower codebook utilization and independent coding rate limit the performance potential of GRs. 

Some other GRs attempt to use finite scalar quantization (FSQ) \cite{2023fsq} or optimized product quantization (OPQ) \cite{2014opq} to tokenize as much effective information as possible \cite{2025rpg}. However, these methods fail to represent core attributes among similar items hierarchically. Therefore, we propose to combine both encoding paradigms. First, we leverage domain knowledge to extract core attributes of queries and items, to enhance the learned semantic and collaborative representations. Then, we use RQ-Kmeans for hierarchical feature encoding and OPQ for quantizing the unique features of each item. This encoding method effectively reduces interference from redundant irrelevant noise, thereby enhancing the relevance between queries and generated items.

\subsubsection{Aligned collaborative and semantic representation}
We integrate semantic knowledge with collaborative signals by aligning the representations of historically interactive query-item pairs. Firstly, we select high-quality query2query, item2item, and query2items pairs from real user search logs using existing retrieval models like ItemCF \cite{2001itemcf} and Swing \cite{2020swing}. Then we collect the content information like query text, item title, item price, keywords, OCR (image-to-text), as well as the statistical business characteristics, such as the number of clicks, add-to-cart, and purchases during a certain time. All these features are processed with a distilled BGE \cite{bge_embedding} to generate a content embedding for each query \(e_q\) and item \(e_i\). Finally, we filter all data with a cosine similarity larger than 0.6 to ensure all pairs are content-relevant.

\begin{table}[tp]  
\caption{18 structured attributes using Named Entity Recognition in the KuaiShou e-commerce search platform.} 
\small
\begin{center}   
\begin{tabular}{|c|c|c|c|c|c|}   
\hline   Entity & Modifier & Brand & Material & Style & Function \\   
\hline   Location & Audience & Color & Marketing & Season & Pattern \\ 
\hline   Scene & Specifications & Price & Model & Anchor &  Series\\ 
\hline
\end{tabular}   
\end{center}   
\label{tab:mostly_words}
\end{table}

We design four types of interrelated tasks to align collaborative and semantic representation: 1) the query2query and item2item contrastive loss \(\mathcal{L_{\text{q2q}}}\), \(\mathcal{L_{\text{i2i}}}\) to align representations of collaboratively similar pairs, 2) a query2item contrastive loss \(\mathcal{L_{\text{q2i}}}\) to ensure that BGE can reflect real business characteristics, 3) a query2item margin loss \(\mathcal{L_{\text{rank}}}\) to further learn the collaborative signal deviation of query-item pairs with different behavior levels (like show, click, order), 4) a hard sample relevance correction loss \(\mathcal{L_{\text{rel}}}\). For pairs with a threshold similarity, we use LLM to score the relevance based on full context information of query and item, and then let the distilled BGE model fit this score.

Then we train the aligned model with the total loss as: 
\begin{equation}
\mathcal{L}_{\text{align}} = \lambda_1 \cdot \mathcal{L_{\text{q2q}}} + \lambda_2 \cdot \mathcal{L_{\text{i2i}}} + \lambda_3 \cdot \mathcal{L_{\text{q2i}}} + \lambda_4 \cdot \mathcal{L_{\text{rank}}} + \lambda_5 \cdot \mathcal{L_{\text{rel}}},
\end{equation}
where \(\lambda_i\) is an adjustable parameter for different objectives.

\subsubsection{Core Keyword Enhancement}
Item textual information often contains redundancy with many irrelevant words and even mutually exclusive attributes. While these attributes can increase item exposure, the weak semantic order caused by numerous stacked attributes makes it difficult for encoders to model key information. Here we propose using core keyword features to enhance textual representation, thereby obtaining semantic IDs dominated by these keywords. 
Specifically, we identified 18 structured attributes (shown in Table \ref{tab:mostly_words}) using Named Entity Recognition (NER) and mined click query-item pairs from the past 1 year as labeled data. We then compiled a list of keywords for each attribute, ranked by page views (PV) in descending order, and selected these high-frequency as core ones. Qwen-VL \cite{Qwen-VL} is employed to identify corresponding keywords for each item, while for queries, we use the Aho-Corasick Automaton \cite{1975ACautomaton} for rapid matching during real-time inference.

All these core keywords are input into the former trained model to obtain vectors \(e_{k}^{i}\) consistent with the item representation distribution. The final optimized representations for each query \(e_{q}^{o}\) and item \(e_{i}^{o}\) are given by:
\begin{equation}
e_{q}^{o} = \frac{1}{2} (e_{q} + \frac{1}{m}\sum_{i=1}^{m}e_{k}^{i}), \quad e_{i}^{o} = \frac{1}{2} (e_{i} + \frac{1}{n}\sum_{j=1}^{n}e_{k}^{j}).
\end{equation}
This approach enhances the role of core keywords in encoding. As shown in Table \ref{tab:Table2}, the core keyword enhancement scheme improves the codebook utilization rate (CUR) of RQ-Kmeans at each level and further increases the independent coding rate (ICR). For example, with a configuration of 4096-1024-512, it results in a 0.10\% CUR increment for Level 1, 24.84\% for Level 2, and 26.15\% for Level 3, as well as the overall ICR increasing by 6.86\%.

\begin{table}[t]
    \caption{The codebook utilization rate (CUR) and independent coding rate (ICR) for various RQ-Kmeans configurations. The last \textsuperscript{+} means balanced operation for all levels.}
    \begin{tabular}{lcccc}
        \toprule
        Configurations & $CUR_{L1}$ & $CUR_{L1*L2}$ & $CUR_{Total}$ & $ICR$ \\
        \midrule
        1024-1024-1024 & 100\% & 54.27\% & 1.72\% & 36.67\% \\ 
        $\backslash$+keywords & 100\% & 65.40\% & 2.03\% & 40.25\% \\
        \midrule
        2048-1024-512 & 100\% & 46.88\% & 1.98\% & 37.80\% \\ 
        $\backslash$+keywords & 100\% & 57.16\% & 2.51\% & 40.76\% \\
        \midrule
        4096-1024-256 & 99.90\% & 39.21\% & 2.27\% & 36.98\% \\ 
        $\backslash$+keywords & 100\% & 48.95\% & 2.94\% & 40.52\% \\
        $\backslash$+l3 balanced & 100\% & 48.95\% & 10.31\% & 60.01\% \\
        \midrule
        4096-1024-512 & 99.90\% & 39.21\% & 1.30\% & 40.54\% \\ 
        $\backslash$+keywords & 100\% & 48.95\% & 1.64\% & 43.32\% \\
        $\backslash$+l3 balanced & 100\% & 48.95\% & 7.03\% & 68.08\% \\
        \midrule
        4096-1024-512\textsuperscript{+} & 99.93\% & 41.45\% & 0.51\% & 33.47\% \\ 
        \bottomrule
    \end{tabular}
    \label{tab:Table2}
\end{table}

\subsubsection{RQ\text{-}OPQ Hierarchical Quantization Tokenization}
\label{RQOPQ}

\begin{table}[t]
    \caption{Performance Comparisons of three Tokenization Schemas. Metrics are evaluated on the real click pairs.}
    \begin{tabular}{lcccc}
        \toprule
        Method & $CUR_{\text{Total}}$ & $ICR$ & Recall@10 & MRR@10 \\
        \midrule
        OnlineMCA & - & - & 0.3440 & 0.1323 \\
        \midrule
        RQ-VAE & 1.17\% & 38.83\% & 0.2171 & 0.0689 \\
        RQ-Kmeans & 7.03\% & 68.08\% & 0.2844 & 0.1038 \\
        RQ\text{-}OPQ & - & \textbf{91.91\%} & \textbf{0.3369} & \textbf{0.1194}\\
        \bottomrule
    \end{tabular}
    \label{tab:Table3}
\end{table}

Common SID tokenizers, such as RQ-VAE, VQ-VAE, and RQ K-means, focus on encoding shared features among similar items, which can result in the loss of distinctive features for each item, ultimately degrading the performance of generative retrieval models (GRs). Furthermore, RQ-VAE has shown weaker performance compared to RQ-Kmeans \cite{2025onerec,ju2025generativerecommendationsemanticids}, as validated in Table~\ref{tab:Table3}. Therefore, we adopt RQ-Kmeans as the foundational tokenizer.

We use the codebook utilization rate (CUR) and the independent coding rate (ICR) as evaluation metrics. The basic codebook size is set to 1024, and the number of codebook layers is set to 3, which aligns with the number of items in the candidate pool. However, e-commerce items have more varied categories and attributes, and RQ-Kmeans tends to prioritize clustering shared prominent features in the former layers. In order to make more concise tokenization, we maintain the capacity of RQ-Kmeans while increasing the codebook size of the former layer to ensure more comprehensive learning of prominent features. As depicted in Table~\ref{tab:Table2}, we tested three configurations: (1024,1024,1024), (2048,1024,512), and (4096,1024,256). The codebook size of 4096 achieves higher CUR and ICR, and the Core Keyword Enhancement scheme ($\backslash$+keywords) shows further improvement. Considering that the search system should encode the entered query similarly and that merchants often increase the number of listed items during global shopping festivals (e.g., 11.11 and 6.18), we further expanded the codebook size to (4096-1024-512). We found that the semantic tokens increased by 11.56\% (as $2 \cdot 1.64\% / 2.94\% -1 $), and the independent coding rate increased to 43.42\% compared to the (4096-1024-512) ($\backslash$+keywords).

To further improve CUR and ICR, OneRec-V1 \cite{2025onerec} proposed using full layers balanced k-means. However, for complex fine-grained attributes of items, forcing them into the same cluster in the early stages can lead to hierarchical clustering collapse. As shown in Table~\ref{tab:Table2}, the $CUR_{total}$ for the balanced k-means operation on full layers (4096-1024-512\textsuperscript{+}) is much lower than the ($\backslash$+keywords) configuration. The $CUR$ drastically decreased from 48.95\% of $CUR_{L1+L2}$ to 1.64\% in $CUR_{total}$, indicating that many similar items were assigned the same ID. Therefore, we propose applying balanced k-means only to the codebook of the third layer to achieve independent encoding of similar items. As shown in ($\backslash$+l3 balanced), the $CUR_{Total}$ increased from 1.64\% to 7.03\%, while the $ICR$ improved by 57.15\%.

Although RQ-Kmeans can construct hierarchical, learnable SIDs for items, it inevitably discards the residual embedding computed after the last clustering. However, this residual embedding contains the distinctive attributes of each item. Therefore, we further use OPQ for quantizing the unique features. The RQ method handles hierarchical semantics, while OPQ is adopted for lateral characteristics. This combined tokenizer can more comprehensively represent the fine-grained features of items, thereby enhancing the relevance constraints for GR models. As shown in Table~\ref{tab:Table3}, the two additional SIDs (256-256) generated by OPQ significantly improve the ICR metric and enhance the recall and ranking capabilities of GRs. More detailed testing is introduced in \S\ref{ablation}.

\subsection{Multi-view Behavior Sequence Injection}
\label{Mu_seq}
We introduce the behavior sequence into GRs from three perspectives. First, we propose a behavior sequence constructed user ID scheme to achieve the distinctive user representation, then explicitly incorporates short behavior sequences in prompt text to learn recent user preferences and implicitly includes long behavior sequences to model user profile, achieving multi-view modeling of user personalized behavior.

\subsubsection{Behavior Sequence Constructed User IDs}
\label{Uid}

Tiger\cite{2023tiger} prepends user-specific tokens into prompts to achieve unique user identification, which is randomly hashed and assigned to a fixed-size vocabulary. However, this method has been shown to be ineffective. We believe that such random IDs do not adequately represent user personalization, as the fixed-size vocabulary may assign the same ID to users with different behaviors. Here, we propose a behavior sequence-constructed user ID scheme to achieve distinctive user representation. Formally, the short behavior sequence consists of the user's latest clicked items, denoted as ${Seq}_{short} = \{s_1, s_2, …, s_m\}$, where $s_i$ is the $i_{th}$ item the user clicked, and $m$ is the total number of latest clicked items. The long behavior sequence is a list of ordered items arranged in chronological order, denoted as ${Seq}_{long} = \{l_1, l_2, …, l_n\}$. The user ID is then computed as the concatenation of $SID_{short}$ and $SID_{long}$:
\begin{equation}
\begin{aligned}
SID_{short} = \lceil \sum_{i=1}^{m} \lambda_i \cdot SID_{s_i} \rceil, \quad \text{where}\ \lambda_i = \frac{exp(\sqrt{i})}{\sum_{i}^{m}exp(\sqrt{i})}, \\
SID_{long} = \lceil \sum_{j=1}^{n} \mu_i \cdot SID_{l_i} \rceil, \quad \text{where}\ \mu_j = \frac{exp(\sqrt{j})}{\sum_{j}^{n}exp(\sqrt{j})}.
\end{aligned}
\end{equation}

So that the length of User IDs is 10. For new-coming or cold-start users, we count the most clicked items for each query based on query-item occurrence and sort them in reverse order by page views as default behavior sequences.

\subsubsection{Explicit Short Behavior Sequence}
\label{sseq}
The short (recent) and long behavior sequences are crucial features in modeling user preferences. The short behavior sequence primarily reflects recent user preferences, while the long behavior sequence represents a user's profile. For example, a soon-to-be-enrolled college student may recently purchase items related to a new dormitory or their major. In contrast, purchases made six months ago might have been more related to college entrance exams or stationery. Therefore, for a user's next search, we prioritize the recent behavioral information. However, the long behavior sequence contains a user's long-term, stable preferences, such as a preference for cost-effectiveness, quality, or style. For the generative retrieval paradigm, explicitly inputting short behavior sequences makes it easier for the model to predict which categories of items users are most likely to click on.

For an e-commerce search platform, the short behavior sequences include the user's latest entered queries $Seq_{query}$ and clicked items $Seq_{short}$. We directly input the SIDs of these queries and items into the prompt, following the constructed user ID and the input query.

\subsubsection{Implicit Long Behavior Sequence}
\label{lseq}
For e-commerce platforms, user long behavior sequences primarily consist of three types: the click, the order, and the search relevant unit (RSU) \cite{guo2023query} sequence. The length of these behavior sequence is almost up to $10^{3}$. Therefore, it is almost impossible to integrate these information into the format of a handcrafted textual prompt. For each item within these sequences, we first map its keyword-enhanced embedding $e_{i}^{o}$ to a corresponding semantic ID (\textbf{\textit{sid}} for simplicity), and then get RQ clustering centroid representation through the lookup method, which is considered to contain multiple levels of semantic information. Furthermore, we aggregate the centroids vector in the same levels, which not only allows the GR to systematically learn the user preferences at different levels, but also saves a lot of resources.

The overall pipeline is as follows, each item is represented with the aggregation of its RQ cluster centroid vectors (3 embeddings):
\begin{align*}
\text{Item}_{sid} &= \text{RQ-Kmeans}(e_i^o), \\
\text{Item}_{emb} &= \text{Emb\_lookup}(\text{Item}_{sid}).
\end{align*}

For long-term historical behavior sequence, the implicit vector representation is computed as
\begin{equation}
\begin{aligned}
\mathbf{M}_{click} = \Bigg\{\sum_{i=1}^{m} \mathbf{Item}_{emb}^{L_1}, \sum_{i=1}^{m} \mathbf{Item}_{emb}^{L_2}, \sum_{i=1}^{m} \mathbf{Item}_{emb}^{L_3} \Bigg\} \\
\mathbf{M}_{order} = \Bigg\{\sum_{i=1}^{n} \mathbf{Item}_{emb}^{L_1}, \sum_{i=1}^{n} \mathbf{Item}_{emb}^{L_2}, \sum_{i=1}^{n} \mathbf{Item}_{emb}^{L_3} \Bigg\} \\
\mathbf{M}_{RSU} = \Bigg\{\sum_{i=1}^{k} \mathbf{Item}_{emb}^{L1}, \sum_{i=1}^{k} \mathbf{Item}_{emb}^{L_2}, \sum_{i=1}^{k} \mathbf{Item}_{emb}^{L_3} \Bigg\} \\
\mathbf{Q} = \text{QFormer}(\mathbf{M}_{click}, \mathbf{M}_{order}, \mathbf{M}_{RSU})
\end{aligned}
\end{equation}
where $\mathbf{M}_{click}$, $\mathbf{M}_{order}$, and $\mathbf{M}_{RSU}$ are referred as the click / order / RSU behavior sequence vectors. $Q$ is the computed long behavior representation. To be specific, $\mathbf{Q} \in \mathbb{R}^{N_M \times d_{model}} \ (d_{model} = 768)$.

These three user behavior sequence injection methods efficiently model users' short-term and long-term personalized preferences from different perspectives. Unlike the stacked behavior sequence concatenation used in MCA ranking stage, this modeling approach not only utilizes resources efficiently but also fully leverages the inference capabilities of GR models.

\subsection{Unified Encoder-Decoder Architecture}
\label{Architecture}
In this section, we introduce the construction of OneSearch from the perspective of feature engineering. 
The input of OneSearch \(\mathbf{X}_{U}\) consists of four parts: 1) User distinctive ID, denoted as \(uid\), which is constructed as detailed in \S~\ref{Mu_seq}. 2) Entered query \(q\), as well as its SID  \({SID}_{q}\); 3) User Short Behavior Sequence, containing the historical search queries \({Seq}_q = \{q_1, q_2, \dots, q_n\}\), the short clicked item sequence \({Seq}_{short} = \{s_1, s_2, \dots, s_n\}\). 4) Implicit long behavior sequence, denoted as \({Seq}_{long}^{emb} = \{l_1, l_2, \dots, l_n\}\). 5) user profile information \(\mathcal{U}\), which is the crowd portrait fitted by the platform. Then OneSearch directly outputs the corresponding item lists \(\mathcal{I}\). OneSearch can adopt either encoder-decoder models (e.g. BART \cite{lewis2019bart}, mT5 \cite{xue2020mt5}), or the decoder-only models (e.g. Qwen3 \cite{yang2025qwen3}) as the backbone \(\mathcal{M}\). The inference flow can be formalized as:
\begin{equation}
\begin{aligned}
\mathcal{I} := \mathcal{M}({uid}, \textit{{q}}, {SID_{q}}, {Seq}_q, {Seq}_{short}, {Seq}_{long}^{emb}, \mathcal{U}).
\end{aligned}
\end{equation}

As illustrated in Figure~\ref{figure4}, our model adheres to the transformer-based \cite{vaswani2017attention} architecture, comprising an encoder that models <\textit{user}, \textit{query}, \textit{behavior sequence}> information, and a decoder dedicated to item generation. We adopted the encoder-decoder models for the real online deployment, as it is effective, have architecturally accelerated training and inference performance. For the unified training, we insert a start token \( t_{\text{[BOS]}} \) and a ending token \( t_{\text{[EOS]}} \) at the first and last place, as well as a separate token \( t_{\text{[SEP]}} \) between adjacent elements to form the input to the encoder. The inference output of \(\mathcal{M}\) is the SIDs, and it can be adjusted throughout constrained or unconstrained beam search.  While constrained beam search guides output to valid SIDs, it increases the decoding complexity (inference time), and unconstrained search explores all sequences without explicit rules. GRID \cite{2025GRID} has shown a similar performance to these two streaming. We also conduct the testing in \S~\ref{ablation}.

\subsection{Preference Aware Reward System}
\label{PARS}
Compared to the sequence coherence in recommendation systems, the strong relevance constraints between queries and items in search engines pose greater challenges for online MCA, often addressed by an independent relevance module in the ranking stage. However, achieving a trade-off between relevance and ranking is a typical Pareto optimality problem. For GR models, it is necessary not only to achieve semantic alignment between SIDs and the textual descriptions of queries and items but also to directly generate items that meet query relevance constraints and user preferences based on historical behavior sequences. The items generated by beam search should naturally balance conversion and correlation. Therefore, we propose a preference aware reward system, which includes a multi-stage supervised fine-tuning (SFT) and an adaptive reward system, to enhance the model's personalized ranking capability. The overall training framework is depicted in Figure~\ref{figure4}(f).

\subsubsection{Multi-stage Supervised Fine-tuning}
\label{Multi-stage SFT}
Considering that the basic architecture (e.g., BART, T5) is pretrained with a large text corpus, but the input queries and items in OneSearch are represented using SIDs, it is essential to first achieve semantic alignment between SIDs and their corresponding textual descriptions. Subsequently, the model should be instructed to generate the desired items that align with user intentions. To address this, we have designed a multi-stage supervised fine-tuning (SFT) procedure. 

\begin{enumerate}
\item \textbf{Semantic Content Alignment}: We set three sub-tasks: (a) Take the query/item text into the prompt as inputs and output the corresponding SIDs. (b) Take the SID as input and generate the original query/item text. (c) Input the query/item text and output the corresponding category information. The first two tasks aim to align the SID and text content, while the category prediction ensures relevance.
    
\item \textbf{Co-occurrence Synchronization}: This stage includes mutual prediction between query and item, and the same task between query SID and item SID. Here, user characteristics are ignored, aiming to learn the intrinsic semantics and collaborative relationships between queries and items based on a large amount of online interactive corpus.

\item \textbf{User Personalization Modeling}: After the aforementioned two stages, we introduce user information into the final stage, which aligns with online inference. Specifically, we concatenate user ID (\S~\ref{Uid}), query, \({{SID}_q}\), \({Seq}_{q}\), \({Seq}_{short}\) (\S~\ref{sseq}), and \({Seq}_{long}^{Emb}\) (\S~\ref{lseq}) as input, with item SID as the training label, to instruct model to learn distinctive personalization.
\end{enumerate}

It should be noted that sliding window data augmentation is applied to the short behavior sequence to guide the model in learning changes in user interests and preferences. The sliding window strategy generates a new segment of the sequence and its subsequent item as the prediction target at each step by sliding a window along the user’s $Seq_{short}$ \cite{zhou2024contrastive}. To prevent the window from becoming too large, we limit the maximum window length. This means we can augment $m$ samples for $Seq_{short} = \{s_1, s_2, …, s_m\}$, with the first sample having no sequence, and the second having the sequence with only one item $s_1$. This approach has been validated for achieving robust and high-performing sequential recommend and generative retrieval models \cite{2025GRID,zhou2024contrastive}. For e-commerce search, this also helps handle new users with limited search history by training on shorter subsequences. More details are shown in \S~\ref{ablation}.

\subsubsection{Adaptive Reward System}
\label{Multi-stage SFT}
Unlike OneRec-V1 \cite{2025OneRecV1}, which uses a weighted P-Score (Preference Score) of multiple objectives to train one reward model followed by Early Clipped GRPO to guide the model in learning user preferences, here we use real online user interactions as feedback signals. These interactions provide more accessible and hierarchical feedback information. 
While this approach shares similarities with the recently proposed OneRec-V2 \cite{zhou2025onerecv2technicalreport}, there are significant differences in training data sampling and training paradigms. We adopt adaptive-weighted reward signals \cite{2025onesug} to construct training data and implement a user-behavior-guided hybrid ranking framework to achieve personalized preference ranking.

\begin{table*}[t]
\centering
\caption{The overall procedure of the preference aware reward system. It contains a three-stage supervised fine-tuning schema for semantic alignment, co-occurrence synchronization, and user personalization modeling, as well as an adaptive reward system for the personalized preference ranking.}
\begin{tabular}{lcccc}
\toprule
\textbf{Procedure} & \textbf{SFT Stage 1} & \textbf{SFT Stage 2} & \textbf{SFT Stage 3} & \textbf{RL Stage} \\
\midrule
\textbf{Objective} 
& Semantic alignment  
& $\langle\text{\(q\), \(i\)}\rangle \ co\text{-}occurrence$
& User personalization
& Preference Alignment \\
\hline
\textbf{Component} 
&$\begin{array}{c}
\text{query} \leftrightarrow \text{SID} \\
\text{item} \leftrightarrow \text{SID}\\
\text{query/item} \mapsto \text{category} \\ 
\text{SID} \mapsto \text{category}
\end{array}$
&$\begin{array}{c}
\text{query} \leftrightarrow \text{item} \\
\text{query\_SID} \leftrightarrow \text{item\_SID}
\end{array}$
& $\begin{bmatrix} uid\ \&\ q\\ {SID}_q\ \& \ Seq_q \\ \text{\(Seq_{short}\)} \\ \text{\(Seq_{long}^{emb}\) } \end{bmatrix} \mapsto \text{item\_SID}$
& $\begin{bmatrix} \text{user \& query} \\ \text{seq. feat.} \\ \text{$item_{win}$} \\ \text{$item_{lose}$} \end{bmatrix} \mapsto \text{Rank Score}$ \\
\bottomrule
\label{table sft}
\end{tabular}
\end{table*}

\textbf{Adaptive-weighted Reward Signal.} Following OneSug \cite{2025onesug} we categorize user interactive behaviors in the search system into six distinct levels: (1)items purchased in search scenario, (2)items of the same category purchased in recommendation scenarios, (3)clicked items, (4) items exposed but not clicked, (5)unshow items in the same category, and (6)random items from other categories. We assign base reward weights as ${\lambda = [2.0, 1.5, 1.0, 0.5, 0.2, 0.0]}$ for each level respectively. Considering that items with higher CTR and CVR in recent days are more likely to be selected by users, we utilize these two metrics to construct adaptive-weighted rewards. However, CTR and CVR often suffer from biased estimation. For example, a newly released item that was exposed only once and then clicked would have CTR at 100\%. Conversely, genuinely popular items are often exposed by Online MCA under various similar but suboptimal queries, resulting in lower CTR and CVR. Therefore, we calibrate these two metrics as follows:
\begin{equation}
\begin{aligned}
Cnt_{T} = \log((Cnt_{pos}+10) \cdot (Cnt_{clk}+10) \cdot (Cnt_{order}+10)) \\
\end{aligned}
\end{equation}
thus:
\begin{equation}
\begin{aligned}
Ctr_i = \frac{\log(Cnt_{clk} + 10)}{Cnt_{T}}, \quad Cvr_i = \frac{\log(Cnt_{order} + 10)}{\log(Cnt_{clk} + 10)}.
\end{aligned}
\end{equation}

The weighted reward score is then defined as:
\begin{equation}
r(q, i) = 2\lambda \cdot \frac{Ctr_i \cdot Cvr_i}{Ctr_i + Cvr_i}.
\end{equation}

For each positive sample $i_{pos}$ and negative sample $i_{neg}$, the user preference difference
$rw_{\Delta}$ is computed as:
\begin{equation}
\begin{aligned}
rw_{\Delta} = \frac{1.0}{r(q, i_{pos}) - r(q, i_{neg})},
\end{aligned}
\end{equation}
where smaller $rw_{\Delta}$ values encourage the model to distinguish nuanced differences in user interactive behaviors.

\textbf{Reward Model Training.} As discussed in OneRec-V2, the reward model in OneRec-V1 employs restricted sampling from a small subset of users to approximate global behavior, potentially learning specific patterns or biases that do not yield actual improvements. However, we also diverge from the feedback-driven preference alignment proposed in OneRec-V2, as the adoption of GRPO and its variants (e.g., ECPO, GBPO) tends to introduce more irrelevant SIDs, and preference rewards require careful tuning for e-commerce search.
Here we design an intuitive and effective reward model based on the Search-based Interest Model (SIM \cite{qi2020searchbasedusermodelinglifelong}) with a three-tower architecture. Each tower is dedicated to learning specific objectives—CTR, CVR, and CTCVR \cite{ma2018entirespacemultitaskmodel} —using binary cross-entropy loss. The final preference score is computed as:
\begin{equation}
\begin{aligned}
Rscore = \lambda_1 \cdot CTR + \lambda_2 \cdot CVR + \lambda_3 \cdot CTCVR + 10 \cdot \lambda_4 \cdot S_{Rel},
\end{aligned}
\end{equation}
where \(\lambda_i\) represents tuned weights (set to 1 in our experiments). To ensure that results generated by OneSearch meet relevance constraints, we additionally incorporate an offline-calculated relevance score $S_{Rel}$ with an amplified weight (10 $\cdot$ $\lambda_4 $).

This reward model differs from the click prediction model in the ranking stage of online MCA in two key aspects: (1) Feature dimensionality: While the ranking model utilizes thousands of features, our reward model only takes user ID, entered query, user behavior sequence, and profile as input, matching OneSearch's input space. (2) Sampling strategy: We additionally include items from the same category clicked in recommendation scenarios as training samples, with labels (1,1,1) for purchased items and (1,0,0) for clicked items. For computational efficiency, the reward model can directly leverage the online MCA ranking model, as we only distill the ranking order rather than absolute scores.

\textbf{Hybird Ranking Framework.}
The alignment stage comprises two components. First, we collect entered queries from real search logs and use the reward model to rerank items output by the fine-tuned OneSearch. We then select samples where ranking changes occur for list-wise DPO training. Items that are clicked or advanced in position by the reward score serve as positive samples, while items pushed back in ranking and those at lower positions serve as negative samples. This allows us to gather one positive sample and multiple negative samples per query for training. The optimization objective can be described as follows:
\begin{equation}
\begin{aligned}
&\mathcal{L} = -\mathbb{E} \Biggl[ \log \sigma \bigg( \log \sum_{i_l \in \mathcal{I}_l} \exp \Big( rw_{\Delta} \\
&\max \big(0, \hat{r}_\theta(x_u, i_w) - \hat{r}_\theta(x_u, i_l) - \delta \big) \Big) \bigg) + {\alpha \log\pi_\theta\left(i_w | x_u \right)} \Biggr], \\
\end{aligned}
\end{equation}
where \( {\mathcal{I}_l} \) denotes the set of negative samples, and \( \hat{r}_\theta(x_u, i_w) \) and \( \hat{r}_\theta(x_u, i_l) \)  represent rewards implicitly defined by the language model \( \pi_\theta \) and reference model \( \pi_{\text{ref}} \):
\begin{equation}
\begin{aligned}
&\hat{r}_\theta(x_u, i_{w/l}) = \beta \log \frac{\pi_\theta(i_{w/l}|x_u)}{\pi_{\text{ref}}(i_{w/l}|x_u)}.
\end{aligned}
\end{equation}

The term $\log\pi_\theta\left(i_w | x_u \right)$ represents the log-likelihood (NLL loss) from the SFT stage. Noted that by combining the list-wise preference alignment with log-likelihood prediction of preferred samples, we establish a novel hybrid paradigm for generative ranking.

This approach primarily trains the initial OneSearch model, as the reward model requires additional computational resources for online generation and scoring. Moreover, since the reward model is trained on user interaction data from the traditional search system, it inherently limits OneSearch's ability to exceed Online MCA's performance ceiling. Therefore, in the second phase, we train the model using pure user interactions. Specifically, we collect positive samples from the first three interactive levels and negative samples from the last three levels, continuing training with the same loss.

In practice, we periodically perform the first RL training using reward model-generated samples. This ensures the model adheres to the online distribution and learns capabilities from the MCA ranking model, which is trained with thousands of features and more parameters. The second phase, preference learning based on user interaction data, is updated as close to streaming as possible. This design aims to overcome online distribution limitations and further leverage the inference capabilities of generative models.

\section{Experiment}
In this section, we conduct comprehensive evaluations on practical industry datasets offline and rigorous A/B online tests to verify the feasibility of OneSearch. Furthermore, we would explore some ablation studies to facilitate further research on a unified end2end generative model for online serving.

\textbf{Datasets} 
We extracted the highly reliable user interactive pairs from Kuaishou's mall search platform between May 2025 and August 2025 to facilitate the supervised fine-tuning (SFT) and DPO. It contains about 1 billion PVs, and all the following offline and ablation experiments were conducted on the full or part of this data. The collections spanned 91 days, with the first 90 days used for model training and the last day used as the test set.

\textbf{Evaluation Metrics} Similar to OneSug, here we take into account the recall and ranking performance. We employed HitRate@K and Mean Reciprocal Ranking (MRR) as the evaluation metrics, which are widely used in search and recommendation systems. All data presented were the average values for all tests.

\textbf{Baseline Methods} To more accurately evaluate the performance of OneSearch, we compare it with the output results of the real online multi-stage cascading architecture (referred to as onlineMCA). Noted that, unlike OneSug, we do not construct an offline MCA system, as the combined models cannot accurately reflect the performance of the online system. Real online e-commerce platforms typically employ multiple recall mechanisms and complex ranking processes with thousands of feature combinations. Using only one model at each stage for simulation would result in an unfair comparison of offline performance.

\textbf{Implementation Details}
We adopt Bart-B \cite{lewis2019bart} as the base pre-trained model for the testing and online deployment, as it is an efficient model with optimized architectural acceleration, and has been online applied in many scenarios in Kuaishou. Due to commercial confidentiality, we do not disclose the total parameters of the online model here, but it is at least 100 times larger than Bart.
The beam search size is set to 512 here to strike a balance between generation quality and latency. The maximum window length is set to n=5. The batch size for SFT and DPO is set to 512 and 128, respectively, with the latter being smaller because the list-wise DPO training takes more samples as inputs. For RQ-OPQ, the number of codebook layers \textit{C} = 5 (3 layers for RQ-Kmeans, and 2 layers for residual OPQ). The codebook size \textit{W} of each layer is (4096,1024,512|256,256). Some of the hyperparameters‌ will be discussed in the following ablation study. The multi-stage supervised training is conducted every week, RL with the reward system is conducted daily, and the hybrid preference alignment with user interaction data is updated as close to the stream as possible. Actually, RL with a reward system can also be trained every week, as we found it does not bring significant performance gains, except during the global shopping festivals (e.g., 11.11 and 6.18).

\subsection{Offline Performance}
We use the real onlineMCA system as the baseline. Specifically, we selected 30,000 pairs of data with click behavior and 30,000 pairs with order behavior from user search logs. We then calculated the Hitrate@350 and MRR@350 for the top 350 items with real user interactions. For a comprehensive evaluation, we also computed the metrics for the data output by pre-ranking stages, but without the final ranking. As shown in Table~\ref{tab:offline_test}, we found that the pre-ranking stage tends to aggregate items with user interactions (resulting in higher recall but much lower MRR), whereas the ranking stage focuses on placing intent items higher in the list. This highlights the optimization objective collision across MCA stages, as the final ranking can only reorder the items output by the pre-ranking stage, ultimately limiting the potential of the final ranking.

\begin{table}[t]
\centering
\caption{Offline performances of our proposed method with onlineMCA on the industry dataset. The best results are in bold, and sub-optimal results are underlined in each column. The "w/o ranking" means "without ranking", and the "$\backslash$+\ keywords " means "add keywords optimizations"}
\begin{tabular}{p{2.0cm}cc|cc}
\toprule
\multirow{2}{*}{\centering Method} & \multicolumn{2}{c|}{order (30k)} & \multicolumn{2}{c}{click (30k)} \\
\cmidrule(lr){2-3} \cmidrule(lr){4-5}
 & \small HR@350 & \small MRR@350 & \small HR@350 & \small MRR@350 \\
\midrule
\small OnlineMCA & 51.74\% & 19.26\% & 64.40\% & 16.89\% \\
\small w/o ranking & 75.75\% & 4.19\% & 80.23\% & 3.00\% \\
\midrule
\small OPQ (8/256) & 19.43\% & 9.55\% & 22.57\% & 7.42\% \\
\small (1024-1024-1024) & 57.39\% & 9.12\% & 63.63\% & 7.46\% \\
\small (2048-1024-512) & 58.29\% & 10.79\% & 65.39\% & 8.86\% \\
\small (4096-1024-256) & 58.57\% & 11.21\% & 64.51\% & 9.24\% \\
\small (4096-1024-512) & 59.58\% & 14.29\% & 62.49\% & 11.82\% \\
\small $\backslash$+\ keywords & 62.38\% & 14.30\% & 66.14\% & 12.10\% \\
\small $\backslash$+\ l3 balanced & 63.16\% & 13.59\% & 68.26\% & 11.67\% \\
\small $\backslash$+\ Adaptive RS & 64.33\% & \underline{16.11\%} & \underline{68.94\%} & \underline{13.80\%} \\
\midrule
\small RQ-OPQ (2/256) & \underline{65.05\%} & 15.33\% & 68.88\% & 12.90\% \\
\small $\backslash$+\ Adaptive RS & \textbf{66.46\%} & \textbf{18.38\%} & \textbf{71.06\%} & \textbf{16.33\%} \\
\bottomrule
\end{tabular}
\label{tab:offline_test}
\end{table}

We tested the RQ-Kmeans and KHQE tokenization with different configurations. As shown in Table~\ref{tab:Table2}, we found that higher codebook utilization rate (CUR) and independent coding rate (ICR) lead to better recall and ranking performance. Additionally, we tested the effects of adding core keyword enhancement, L3 balanced kmeans, and the adaptive reward system. All of these enhancements improved the metrics to varying degrees. Notably, the adaptive reward preference learning significantly improved the model's ranking capability, with average improvements of 1.80\% and 3.24\% in HR@350 and MRR@350, respectively.

The final solution, described as RQ-OPQ (2/256) $\backslash$+\ Adaptive RS, means the model adopts the tokenization of RQ-Kmeans (4096-1024-512) followed by OPQ (256-256), and trained by the full preference aware reward system. This configuration achieved a much higher recall metric (66.46\% vs. 51.74\% for order) and comparable MRR performance (18.38\% vs. 19.26\% for order) compared to the onlineMCA. We believe this approach can ensure personalized ranking capability while maximizing the placement of items that match the search intent at the front of the list. This configuration would be called OneSearch in the following section for brevity.

\begin{table}[t]
\centering
\caption{Ablation study of multi-view behavior sequence injection. Slid. Window means the sliding window strategy.}
\begin{tabular}{p{2.0cm}cc|cc}
\toprule
\multirow{2}{*}{\centering Method} & \multicolumn{2}{c|}{order (30k)} & \multicolumn{2}{c}{click (30k)} \\
\cmidrule(lr){2-3} \cmidrule(lr){4-5}
 & \small HR@350 & \small MRR@350 & \small HR@350 & \small MRR@350 \\
\midrule
\small OneSearch & 66.46\% & 18.38\% & 71.06\% & 16.33\% \\
\midrule
\small w/o User SIDs & -0.94\% & -0.37\% & -1.72\% & -0.36\% \\
\small w/o $Seq_{short}$ & -3.43\% & -1.53\% & -4.15\% & -1.32\% \\
\small  w/o $Seq_{long}^{emb}$ & -2.26\% & -1.01\% & -3.00\% & -1.05\% \\
\small w/o Slid.Window & -1.95\% & -0.81\% & -1.80\% & -0.70\% \\
\bottomrule
\end{tabular}
\label{tab:ablation_study4seq}
\end{table}

\subsection{Ablation Study}
\label{ablation}
To further demonstrate the performance of the proposed method, we evaluated the effectiveness of 1) the multi-view behavior sequence injection schema, 2) RQ-OPQ tokenization, showing changes in CUR and ICR over time with the introduction of new items, and 3) different OPQ encoding tokenizations.

The first evaluation is of different behavior sequences. As shown in Table~\ref{tab:ablation_study4seq}, "w/o User SIDs" replaces the sequence-constructed user IDs (presented in \S~\ref{KHQE}) with Hashing User ID \cite{2023tiger}, resulting in an average decrease of 1.33\% in HR@350 and 0.36\% in MRR. This indicates that constructing User IDs using sequences more adequately represents user personalization compared to assigning a unique ID without semantic and collaborative information. The hashing method has also been shown to have limited improvement in recommendation systems \cite{2025GRID}. We also found that explicitly incorporating short behavior sequences in prompt text and implicitly including long behavior sequences into the model can significantly enhance model performance. Particularly, the short behavior sequence can bring an average increase of 3.79\% in HR@350 and 1.43\% in MRR@350. The sliding window augmentation on the short sequence is also validated to be effective in guiding the model to learn changes in user interests and preferences.

The items in a search system are constantly changing, especially during global shopping festivals, where poorly selling items are removed, and many new items in potentially high-demand categories are introduced. Consequently, a pre-calculated item SIDs pool with balanced k-means is continuously disrupted. Over time, more items would aggregate under the same SIDs. We conducted an experiment to verify whether this change is significant. We used all available items as of July 15 to construct two tokenizers (RQ-Kmeans and RQ-OPQ) and tested the trends in CUR and ICR as new items were added. As shown in Figure~\ref{fig:ICR_SIDratio}, the change in numerical values is minimal, even after the promotions on August 18. For example, RQ-Kmeans saw a 1.11\% decrease in CUR, while RQ-OPQ only decreased by 0.43\%. These results also validate the superiority of RQ-OPQ Tokenization.

\begin{figure}[t]
  \centering
  \includegraphics[width=\linewidth]{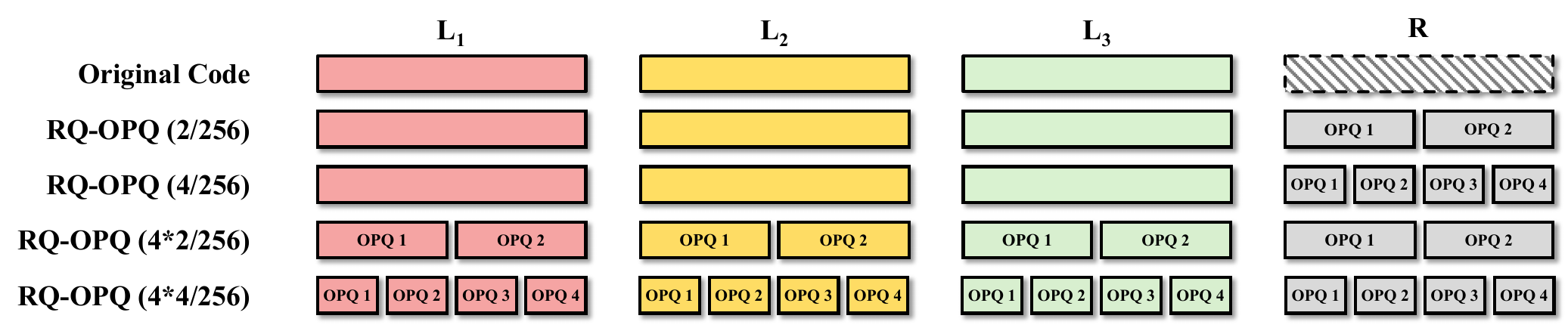}
  \caption{The different hierarchical quantization encodings of items.}
  \label{fig:opq-ablation}
\end{figure}

\begin{figure}[htbp]
\centering
\begin{minipage}[ht]{0.49\columnwidth}
    \includegraphics[width=\columnwidth]{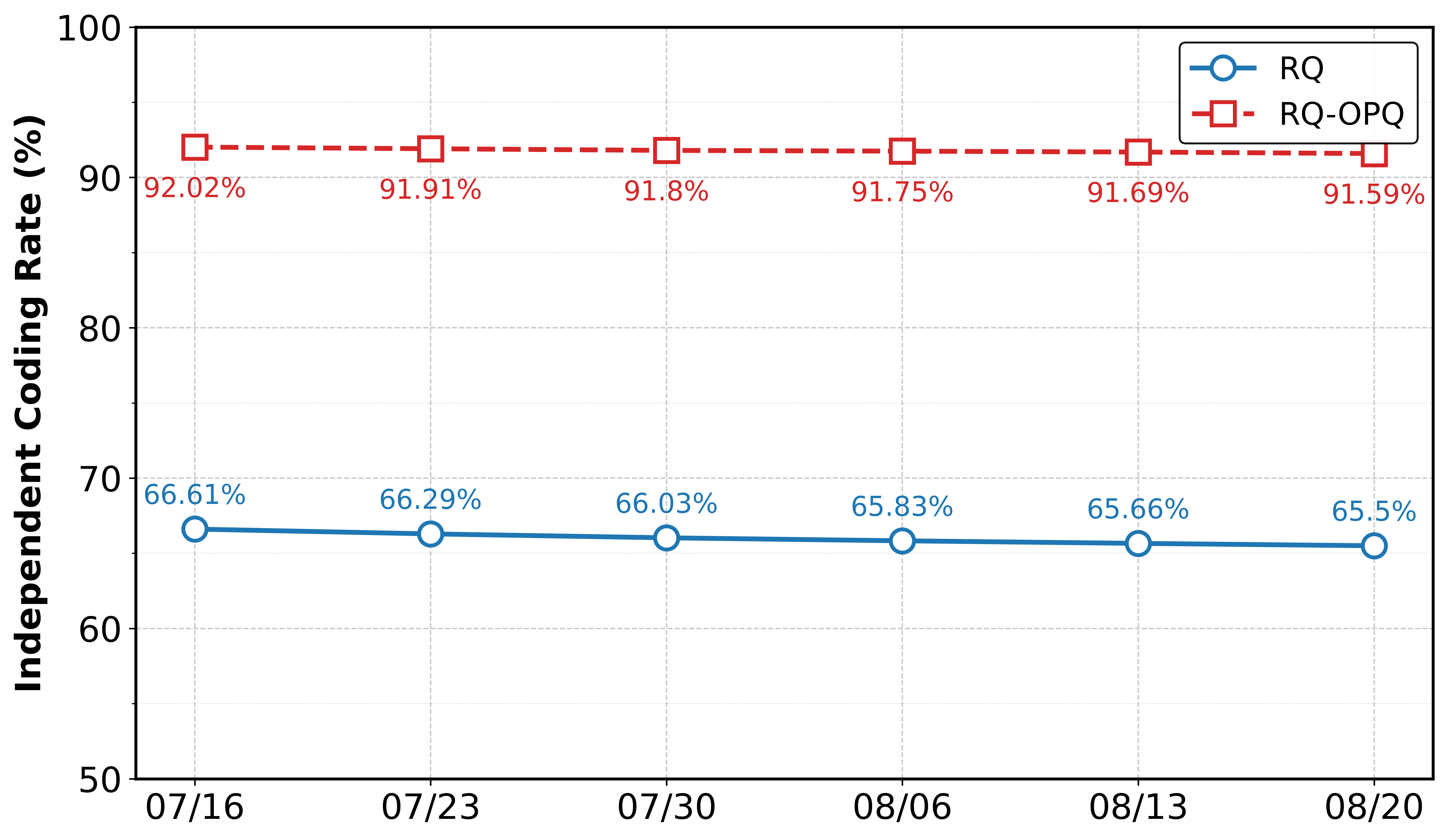}
    \subcaption{ICR}
    \label{fig:ICR}
\end{minipage}
\begin{minipage}[ht]{0.49\columnwidth}
    \includegraphics[width=\columnwidth]{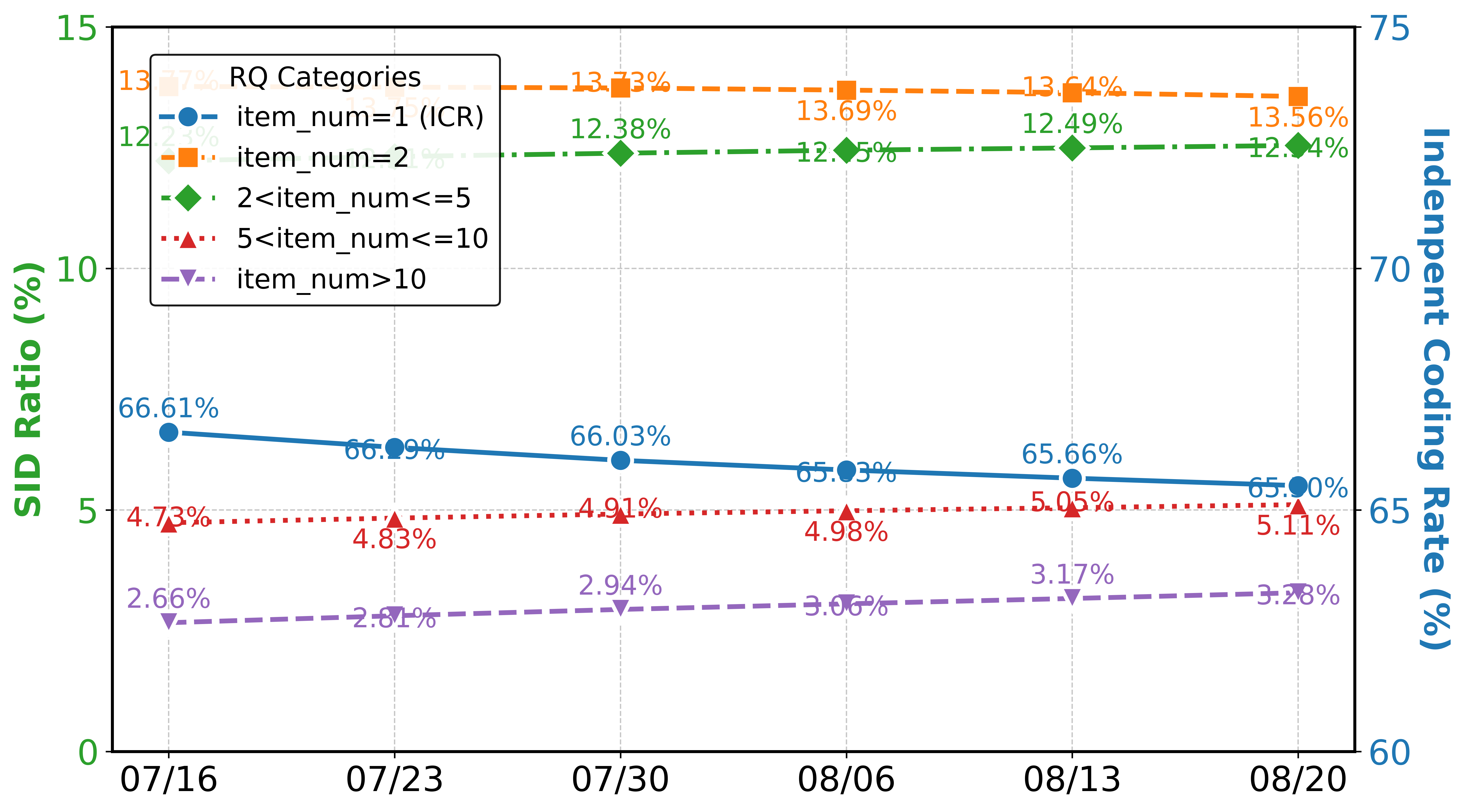}
    \subcaption{SID Ratio}
    \label{fig:SIDratio}
\end{minipage}
\caption{The ICR and SID ratio indicators of RQ-Kmeans after regular time intervals.}
\label{fig:ICR_SIDratio}
\end{figure}

We also examined the impact of different hierarchical quantization encodings on items in Figure~\ref{fig:opq-ablation}. As shown in Table~\ref{tab:ablation_study4opq}, we computed two metrics with the top 10 items for quick validation. RQ-OPQ (2/256) is the basic configuration, and RQ-OPQ (4/256) means the residual embedding is tokenized by OPQ (256-256-256-256). RQ-OPQ (4*2/256) means all embeddings (the cluster of three layers and the residual one) are tokenized with OPQ (2/256), then (4*4/256) indicates further quantization. We found that the basic RQ-OPQ (2/256) achieved the highest performance. (4/256) perform weakly with increased sequence length and decoding complexity. The other two configurations were almost entirely ineffective, which is similar to the balanced k-means operation on full layers in \S~\ref{KHQE}, as the hierarchical features were not distinctly represented, leading to many items being aggregated under the same SID.

\begin{table}[t]
\centering
\caption{Ablation study of different OPQ tokenizations.}
\begin{tabular}{p{2.0cm}cc|cc}
\toprule
\multirow{2}{*}{\centering Method} & \multicolumn{2}{c|}{order (30k)} & \multicolumn{2}{c}{click (30k)} \\
\cmidrule(lr){2-3} \cmidrule(lr){4-5}
 & \small HR@10 & \small MRR@10 & \small HR@10 & \small MRR@10 \\
\midrule
\small RQ-OPQ (2/256) & 28.42\% & 14.15\% & 33.69\% & 11.94\% \\
\midrule
\small *-OPQ (4/256) & -2.36\% & -1.77\% & -2.52\% & -1.56\% \\
\small *-OPQ (4*2/256) & -10.20\% & -5.57\% & -11.77\% & -3.84\% \\
\small *-OPQ (4*4/256) & -24.18\% & -11.83\% & -27.11\% & -9.61\% \\
\bottomrule
\end{tabular}
\label{tab:ablation_study4opq}
\end{table}

\begin{figure}[!thp]
  \centering
  \includegraphics[width=\linewidth]{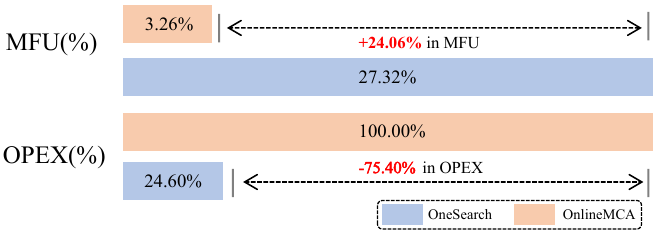}
  \caption{The comparisons of MFU and OPEX for onlineMCA and OneSearch.}
  \Description{query suggestions.}
  \label{fig:MFU}
\end{figure}

\subsection{Online A/B Testing}
\label{ABTEST}
To verify OneSearch’s effectiveness in online applications, we compared it with onlineMCA in KuaiShou's mall search platform through rigorous online A/B tests. It takes the short query the user entered as input and directly outputs the item candidates, where an item with a higher score would be displayed more prominently.
We trained two versions of the OneSearch model successively. $OneSearch^{1}$ refers to the model encoded using RQ-Kmeans and trained without incorporating the implicit long behavior sequence for modeling the user profile. 
$OneSearch^{2}$ is the model with all optimizations. Similar to OneRec-V1 \cite{2025OneRecV1}, we established two experimental groups: one employing a pure generative model ($OneSearch$) and another reordering generative outputs with a reward model based selection ($OneSearch_{RM}$).
As indicated in Table~\ref{tab:online}, the pure generative model with multi-stage supervised fine-tuning and adaptive preference learning can achieve comparable performance to the entire complex search system. By the introduction of RQ-OPQ and long behavior sequence, $OneSearch^{2}$ can confidently improve item CTR by 1.45\%, PV CTR by 1.40\%. Further applying reward model selection ($OneSearch_{RM}^{2}$) achieved statistically significant improvements on all metrics, with 1.67\% in item CTR, 3.14\% in PV CTR, 1.78\% in PV CVR, 2.40\% in Buyer volume, and 3.22\% in Order volume. 
For a clearer comparison, we perform additional experiments on the online search system, named MCA w/o ranking, which only uses the "recall and pre-ranking" module to predict the items, without the ranking stage. It significantly reduces all indicators, especially with 28.78\% in Buyer, 39.14\% in Order volume. This indirectly verifies that OneSearch has comparable ranking capabilities.
These outstanding results show that OneSearch outperforms the onlineMCA, and indicate it can update the complicated online system to a more balanced state without generating seesaw effects.

\begin{table}[!t]
\centering
\footnotesize
\caption{Online results for A/B testing. The black fonts indicate that the statistical significance (P-value) is smaller than 0.05, while the gray ones are larger than 0.05, which means the data are not yet confident.}
\begin{tabular*}{1\linewidth}{@{\extracolsep{\fill}}cccccc@{\extracolsep{\fill}}}
\toprule
\textbf{Method} & \textbf{Item CTR} & \textbf{PV CTR} & \textbf{PV CVR} & \textbf{Buyer} & \textbf{Order} \\
\midrule
MCA w/o ranking & -9.97\% & -20.33\% & -11.55\% & -28.78\% & -39.14\% \\
\midrule
$OneSearch^{1}$ & \textcolor{gray}{-1.10\%} & \textcolor{gray}{-2.06\%} & \textcolor{gray}{+0.39\%} & \textcolor{gray}{+1.27\%} & \textcolor{gray}{-2.22\%} \\
$OneSearch_{RM}^{1}$ & +1.40\% & +3.05\% & +1.94\% & +1.92\% & +1.59\% \\
\midrule
$OneSearch^{2}$ & +1.45\% & +1.40\% & \textcolor{gray}{-0.12\%} & \textcolor{gray}{-0.58\%} & \textcolor{gray}{-0.69\%} \\
$OneSearch_{RM}^{2}$ & \textbf{+1.67\%} & \textbf{+3.14\%} & \textbf{+1.78\%} & \textbf{+2.40\%} & \textbf{+3.22\%} \\
\bottomrule
\end{tabular*}
\label{tab:online}
\end{table}

We also measured Model FLOPs Utilization (MFU) on flagship GPUs during serving inference. In Figure~\ref{fig:MFU}, the onlineMCA is only 3.26\%, but OneSearch can achieve 27.32\%, with a relative improvement of 700.38\%. This significantly outperforms onlineMCA and the common large language models (LLMs), which typically reach ~40\% of MFU on H100 GPUs \cite{dubey2024llama}. 
Furthermore, OneSearch significantly reduces communication and memory overhead, resulting in operational expenditure (OPEX) reduced to only 24.60\% of the online search pipeline, promoting the application of GRs in search systems.

Last but not least, to ascertain the actual impacts on the online search experience, we conducted additional manual evaluations. We randomly selected 200 queries and extracted 3,200 query-item pairs from identical exposure positions, ensuring all other variables remained constant. We set three metrics as 1) page good rate - an evaluation indicator for the overall user experience, 2) item quality - Check whether the displayed products are counterfeit, have mismatched images and text, or have abnormal prices, and 3) query-item relevance - we engaged experts to rate each pair as "Good" (both subject and core keywords match), "Fair" (only subject matches), or "Bad" (subjects differ). The outcomes of these assessments are presented in Table~\ref{tab:manual_evaluation}. We can see that $OneSearch^{2}$ achieves substantial increases in page good rate by 1.03\%, item quality by 2.12\%, and query item relevance by 1.87\%. The deployment of RQ-OPQ further enhances the relevance of model generation.

Ultimately, OneSearch has been successfully deployed for the entire traffic on the e-commerce detail page search engine in Kuaishou, 50\% traffic on the mall search, and 20\% traffic on the homepage e-commerce search platform for further investigation, which serves millions of users generating tens of millions of PVs daily.

\begin{table}[t]
\centering
\caption{Manual evaluation results for online experience.}
\begin{tabular}{lccc}
\toprule
Metric & Page Good Rate & Item Quality & Q-I Relevance \\
\midrule
$OneSearch^{1}$ & 0.84\% & 1.69\% & 1.40\% \\
$OneSearch^{2}$ & 1.03\% & 2.12\% & 1.87\% \\
\bottomrule
\end{tabular}
\label{tab:manual_evaluation}
\end{table}

\begin{table}[tp]
\centering
\caption{Online CTR gains for three query popularity.}
\begin{tabular*}{\linewidth}{@{\extracolsep{\fill}}cccc@{}}
\toprule
\textbf{Method} & \textbf{Top} & \textbf{Middle} & \textbf{Long-tail} \\
\midrule
$OneSearch^{2}$ & +1.25\% & +2.27\% & +1.33\% \\
\bottomrule
\label{tab:top_middle_tail_query}
\end{tabular*}
\end{table}

\subsection{Further Analysis}

\begin{figure}[t]
  \centering
  \includegraphics[width=\linewidth]{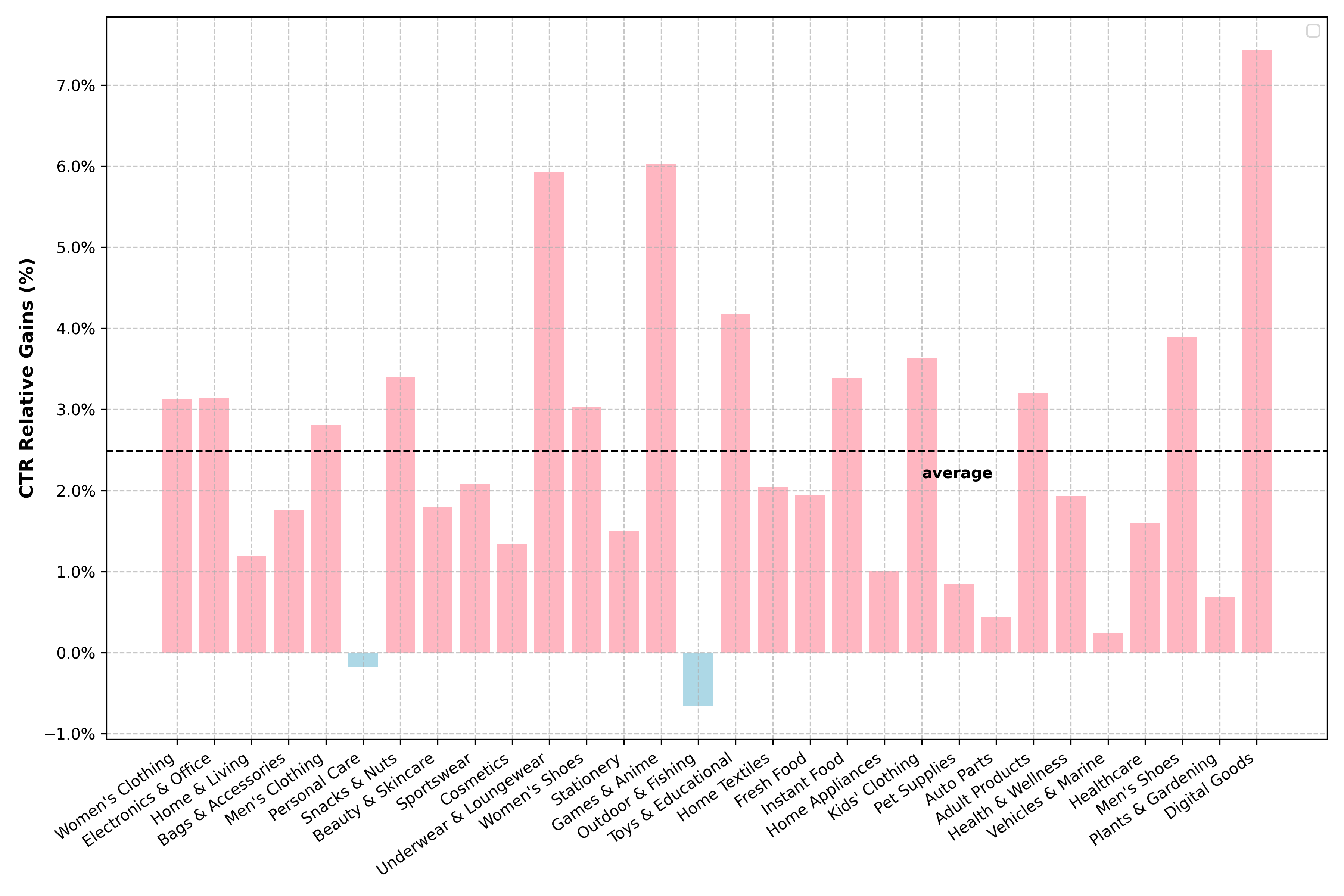}
  \caption{The online CTR relative gains for top 30 industries.}
  \Description{query suggestions.}
  \label{fig:top30_industry}
\end{figure}

In this section, we mainly discuss three questions about the online deployment of the end2end generative framework and provide our investigations to facilitate further research.

1) \textbf{What are the main aspects of the online gains for the OneSearch model?} In our analysis, we focused on the dimensions of industry and query popularity. As illustrated in Figure~\ref{fig:top30_industry}, we calculated the CTR relative gains across the top 30 industries. Remarkably, 28 out of 30 industries experienced increases, with an average gain of 2.49\%. These results were statistically significant, with P-values below 0.05. Although two industries showed negative effects, these were not statistically significant. Overall, the unified modeling optimization demonstrates substantial potential in addressing the inconsistent objectives of multi-stage processes in MCA systems, benefiting nearly all industries.

As for the query popularity dimension, we divided all prefixes into three categories: top (PV number daily larger than 1,000), middle (larger than 100 and less than 1,000), and long-tail (less than 100). The item CTR relative gains for each were listed in Table~\ref{tab:top_middle_tail_query}. Queries of all categories are enhanced with the OneSearch models. These results indicate that the rich semantic and interactive representations induced by keyword-enhanced hierarchical quantization encoding, multi-view behavior sequence, and the preference aware reward system can greatly improve the recognition of e-commerce search for queries of all popularity.

2) \textbf{Does OneSearch have stronger reasoning capabilities?} In traditional e-commerce search scenarios, ranking models often involve thousands of features, and the combination of them can obscure some key attributes. Additionally, the structure of common ranking model typically consists of a simple stack of shallow neural networks, resulting in minimal reasoning capabilities. OneSearch, on the other hand, leverages users' long- and short-term sequential information to identify their potential interests and enhances the inference of user search intent through the attention mechanism of transformer structures. For instance, a female user who previously searched for "couple sneakers" and "Valentine's Day gifts" is likely seeking a pair of rings for both her partner and herself when searching for "silver ring." We observed in real logs that only OneSearch presented the relevant product, which was ultimately purchased by the user.

\begin{table}[tp]
\centering
\caption{Online CTR gains for cold-start items and users.}
\begin{tabular*}{\linewidth}{@{\extracolsep{\fill}}cccc@{}}
\toprule
\textbf{Object} & \textbf{Warm} & \textbf{Cold} & \textbf{Average} \\
\midrule
Item & +2.34\% & +3.31\% & +2.52\% \\
User & +1.11\% & +2.50\% & +2.41\% \\
\bottomrule
\label{tab:cold_start}
\end{tabular*}
\end{table}

3) \textbf{How does OneSearch perform for cold-start users and items?} We conducted tests to evaluate the model's performance in cold-start scenarios. Here, we define cold items as those published within the last seven days with no interaction behavior, and cold users as those who have not used the Kuaishou app in the past 90 days. The specific comparison results are demonstrated in Table~\ref{tab:cold_start}. Compared to the onlineMCA, we found that OneSearch's performance for cold-start items and users has improved by 3.31\% and 2.50\%, respectively. Both of them are greater than the metrics for warm ones. These results show that OneSearch can handle the cold-start issue well.

4) \textbf{What optimization points will OneSearch consider in the future?} The addition of OPQ-based tokenization can even quickly process new hotwords. We constructed a new keyword offline and added it to the textual descriptions of some items. Without reconstructing a new codebook, OneSearch was still able to generate SIDs for these items during inference. This finding further motivates us to consider online real-time encoding. We will explore in future research, aiming to achieve unified encoding and inference using a single generative model, thereby reducing the gap between scheduled encoding and streaming training phrase. Additionally, aligning user preferences through more robust reinforcement learning and incorporating multi-modal features (such as images and videos) for items can further enhance the reasoning capabilities.

\section{Conclusion}
In this paper, we present OneSearch, a pioneering end-to-end generative framework for e-commerce query search that effectively overcomes the limitations of traditional multi-stage cascading architecture. By employing a unified generative model, introducing the keyword-enhanced hierarchical quantization encoding, and injecting the multi-view behavior sequences, OneSearch achieves superior semantic understanding and personalization modeling. The preference aware reward strategy further refines the model's ability to capture user preferences, leading to improved ranking performance. Extensive offline and online evaluations confirm OneSearch's effectiveness in boosting query diversity, click-through rates, and business conversions. Its successful deployment on multiple Kuaishou search scenes underscores its practical applicability and potential to enhance industry revenue. OneSearch sets a new benchmark for industrial query search solutions, paving the way for future advancements in generative retrieval methods.

\bibliographystyle{ACM-Reference-Format}
\bibliography{sample-base}

\end{document}